# A Fuzzy Co-Clustering approach for Clickstream Data Pattern

R.Rathipriya[1] Dr. K.Thangavel[2]

*Abstract*-Web Usage mining is a very important tool to extract the hidden business intelligence data from large databases. The extracted information provides the organizations with the ability to produce results more effectively to improve their businesses and increasing of sales. Co-clustering is a powerful bipartition technique which identifies group of users associated to group of web pages. These associations are quantified to reveal the users' interest in the different web pages' clusters. In this paper, Fuzzy Co-Clustering algorithm is proposed for clickstream data to identify the subset of users of similar navigational behavior /interest over a subset of web pages of a website. Targeting the users group for various promotional activities is an important aspect of marketing practices. Experiments are conducted on real dataset to prove the efficiency of proposed algorithm. The results and findings of this algorithm could be used to enhance the marketing strategy for directing marketing, advertisements for web based businesses and so on.

*Keywords*-Web usage mining, Fuzzy Co-Clustering, Target marketing, Clickstream data

## I. Introduction

Nowdays, internet is a very fast communication media between business organizations' services and their customers with very low cost. Web Data mining [1] is an intelligent data mining technique to analyze web data. It includes web content data, web structure data and web usage data. Analysis of usage data provides the organizations with the required information to improve their performances.  In general, Web clustering techniques are used to discover the group of users or group of pages  called clusters which are similar between them and dissimilar to the users /pages in the other cluster. User clustering approaches of usage data create groups with similar browsing pattern. Web page's content data, structure data and usage data are used to cluster the web pages of a web site. Clustering results may be beneficial for wide range of application such as web site personalization, system improvement, web caching and pre-fetching, recommendation system, design of collaborative filtering and target marketing. These clustering techniques are one dimensional where as Co-Clustering is the bi-dimensional clustering technique. The combination of user cluster with set of its significant web pages of a web site is called a Co-Cluster.

There are many applications for Co-clustering[7] such as recommendations systems, direct marketing, text mining identifying the web communities and election analysis [1][2]. Co-Clustering techniques can be used in, collaborative filtering to identify subgroups of customers with similar preferences or behaviors towards a subset of products with the goal of performing target marketing. Recommendation systems and Marketing are important applications in E-commerce area. The main goal for the above applications is to identify group of web users or customers with similar behavior/ interest so that one can predict the customer's interest and make proper recommendations to improve their sale Generally, Co-clustering is a form of two-way clustering in which both dimensions are clustered simultaneously Target and generated Co-Clusters are refined using some techniques like fuzzy approach.  The goal of this paper is to provide fuzzy Co-Clustering algorithm for clickstream data to quantify the discovered Co-Clusters .Users' clusters with their members are related in a different degree with pages' clusters. The relation between these clusters is quantified using fuzzy membership function to show the distribution of users' interest over the web page clusters.

The organization of the paper is as follows. Section 2 summarizes some of the existing web clustering techniques and co-clustering approaches. Section 3 describes the problem statements. The proposed Fuzzy Co-clustering algorithm is described briefly in Section 4. The experimental results of the proposed algorithm are discussed in the Section 5. Section 6 concludes this paper.

## II. Background

### A. Related work

Web mining was first proposed by Etzioni in 1996.Web mining techniques automatically discover and extract information from World Wide Web documents and services. Cooley et al.[1,6] did in-depth  research in web usage mining. Approaches proposed in [3,10] extend the one dimensional clustering problem and focus on the simultaneous grouping of both users and web pages by exploiting their relations. Its goal is to identify groups of related web users and pages, which has similar interest across the same subset of pages. This behavior reveals users' interests as similar and highly related to the topic that the specific set of pages involves. The obtained results are particularly useful for applications such as e-commerce and recommendation engines, since relations between clients and products may be revealed. These relations are more meaningful than the one dimensional clustering of users or pages.

Co-Clustering algorithms fall into three categories.  First category models each type of object as a random variable, clustering the objects of different types simultaneously while preserving the mutual information between the

About-[1]Department of Computer Science,  Periyar University, Salem,Tamilnadu,India.
(e-mail;[1]rathi_priyar@yahoo.co.in ,[2]drktvelu@yahoo.com)

random variables that model these objects. The second category models the relationship between different types of objects as a (nonnegative) matrix. This matrix is *approximately* decomposed into several matrices, which indicate the cluster memberships for the objects . The third category treats the relationship between different types of objects as a graph and performs co-clustering by graph partitioning based on spectral analysis [2].Fuzzy biclustering approach to correlate the web users and web pages based on spectral clustering technique was proposed in[2].

### B. Co-Clustering Approach

By definition Co-Clustering[9] is the process of simultaneous categorization of user and web pages into user cluster and page cluster respectively. The term co-cluster refers to each pair of user cluster and page cluster. Using the matrix illustration, a co-cluster is represented by a sub-matrix of A where the aij values of all its elements are similar to one another. Thus co-clustering is the task of finding these coherent sub-matrices of A. One illustration of co-clustering is shown in the following matrix. The six square matrices represent the six co-clusters (i.e. A11 to A32).

$$A = \begin{pmatrix} 5 & 5 & 5 & 0 & 0 & 0 \\ 2 & 2 & 2 & 0 & 0 & 0 \\ 0 & 0 & 0 & 1 & 5 & 7 \\ 0 & 0 & 0 & 1 & 5 & 7 \\ 4 & 4 & 0 & 4 & 4 & 4 \\ 4 & 4 & 0 & 4 & 4 & 4 \end{pmatrix} = \begin{pmatrix} A_{11} & A_{12} \\ A_{21} & A_{22} \\ A_{31} & A_{32} \end{pmatrix}$$

This paper aims to provide a framework for the simultaneous clustering of web pages and users called Fuzzy Co-Clustering. The relations between web users and pages in a co-cluster will be identified and quantified. Here, users grouped in the same users' cluster may be related to more than one web pages' cluster with different degree of fuzzy membership value and vice versa.

### III. PROBLEM STATEMENT

This section gives the formal definitions of the problem and describes how the clickstream data from web server log file converted into matrix form.

Let A( U , P) be an 'n x m' user associated matrix where U={$U_1,U_2$ ,……,$U_n$} be a set of users and P={ $P_1,P_2$,……,$P_m$} be a set of pages of a web site. It is used to describe the relationship between web pages and users who access these web pages. Let 'n' be the number of web user and 'm' be the number of web pages. The element aij of A(U,P) represents frequency of the user Ui of U visit the page Pj of P during a given period of time

$$a_{ij} = \begin{cases} Hits(U_i, P_j), & \text{if } P_j \text{ is visited by } U_i \\ 0, & \text{otherwise} \end{cases} \quad (1)$$

where Hits($U_i$ ,$P_j$) is the count/frequency of the user $U_i$ accesses the page $P_j$ during a given period of time.

A **user cluster** is a group of users that have similar behavior when navigating through a web site during a given period of time. A **page cluster** is a group of web pages that are related according to user's perception, specifically they are accessed by similar users during a given period of time. Similarity measure used in this paper is **Fuzzy similarity**. Fuzzy similarity measure between two fuzzy subsets $X_1=\{x_{11},x_{12},\ldots,x_{1n}\}$ and $X_2=\{x_{21},xu_{22},\ldots,x_{2n}\}$ is defined as

$$fsim(X_1, X_2) = \frac{\sum_{k=1}^{m} x_{1k} \wedge x_{2k}}{\sum_{k=1}^{m} x_{1k} \vee x_{2k}} \quad (2)$$

This ratio defines the similarity between two fuzzy subsets, with values between 0 and 1. Using this similarity measure, compute similarity matrix for user vector and page vector of user associated matrix A.

Fuzzy co-clustering is a technique that performs simultaneous clustering of objects and features using fuzzy membership function to correlate their relations. It allows user clusters to belong to several page clusters simultaneously with different degree of membership value. The membership value lies between 0 and 1.

### A. Clustering algorithm : K-Means

In this paper, K-Means clustering technique[12] is used to create user cluster and page cluster. K-means is one of the simplest unsupervised learning algorithms for clustering problem. The procedure is simple and easy way to classify a given data set through a certain fixed number of clusters (assume K clusters).

The algorithm is composed of the following steps:

> Place K points into the space represented by the objects that are being clustered. These points represent initial group centroids.
>
> Assign each object to the group that has the closest centroid.
>
> When all objects have been assigned, recalculate the positions of the K centroids.
>
> Repeat Steps 2 and 3 until the centroids no longer move. This produces a separation of the objects into groups from which the metric to be minimized can be calculated.

The K-Means algorithm is significantly sensitive to the initial randomly selected cluster centers. Run K-Means algorithm repeatedly with different random cluster centers(called centriods)approximately for ten times. Choose the best centriod whose Davis Bouldin Index value is minimum.

### B. Web Server Log File

Web server log file[3,5] is a log file automatically created and maintained by server of activity performed by it. Default log file format is Common log File format. It contains information about the request, client IP address, request date/time, page requested, HTTP code, bytes served, user agent, and referrer. These data can be combined into a single file, or separated into distinct logs, such as an access log, error log, or referrer log. From web server log file, which user access which web page of a web site during a specified period of time can be obtained easily.

### C. Clickstream Data

Clickstream data[4] is a natural by-product of a user accessing world wide web pages, and refers to the sequence of pages visited and the time these pages were viewed. Clickstream data is to Internet marketers and advertisers. An instance of real clickstream records is the MSNBC dataset, which describes the page visits of users who visited msnbc.com on a single day. There are 989,818 users and only 17 distinct items, because these items are recorded at the level of URL category, not at
page level, which greatly reduces the dimensionality. The 17 categories are tabulated with their category number.

| Frontpage | 1 | News | 2 |
|---|---|---|---|
| Tech | 3 | Local | 4 |
| Opinion | 5 | On-air | 6 |
| Misc | 7 | Weather | 8 |
| Health | 9 | Living | 10 |
| Business | 11 | Sports | 12 |
| Summary | 13 | Bbs | 14 |
| Travel | 15 | msn-news | 16 |
| msn-sports | 17 | | |

**Sample Sequences**

1 1
2
3 2 2 4 2 2 2 3 3
6 7 7 7 6 6 8 8 8 8
6 9 4 4 4 10 3 10 5 10 4 4 4

Each row describes the hits of a single user. For example, the first user hits "frontpage" twice, and the second user hits "news" once.

### D. User fuzzy subset

For each user $U_i$, use the user accessing information on each web page $P_j$ to describe the visiting pattern. Then user fuzzy subset $\mu_{U_i}$ of $i^{th}$ user that reflects the user's visiting behavior is defined as

$$\mu_{U_i} = \{(P_j, f\mu_{U_i}(P_j)) \mid P_j \in P\}$$

where $f\mu_{U_i}(P_j)$ is the membership function which is defined as

$$f\mu_{U_i}(P_j) = \frac{\text{Hits}(U_i, P_j)}{\sum_{k=1}^{m} \text{Hits}(U_i, P_k)} \quad (3)$$

and m is the number of web pages of a web site.

### E. Page Fuzzy subset

For each page $P_j$, use the all user accessing information on the web page $P_j$ to describe the web page itself. Then page fuzzy subset $\mu P_j$ that reflects all users' visiting behavior on the jth page is defined as

$$\mu_{P_j} = \{(U_i, f\mu_{P_j}(U_i)) \mid U_i \in U\}$$

where $f\mu_{P_j}(U_i)$ is the membership function which is defined as

$$f\mu_{P_j}(U_i) = \frac{\text{Hits}(U_i, P_j)}{\sum_{k=1}^{n} \text{Hits}(U_k, P_j)} \quad (4)$$

and n is the number of web users.

### IV. FUZZY CO-CLUSTERING ALGORITHM FOR CLICKSTREAM DATA

In this paper, K-Means clustering method is applied on the user(row) and page(column) dimensions of the user access matrix A(U,P) separately and, then combine the results to obtain small co-regulated submatrices called Co-Clusters. Given a user access matrix A, let ku be the number of clusters on user dimension and kp be the number of clusters on page dimension after K-Means clustering is applied. Cu is the family of user clusters and Cp is the family of page clusters. Let ciu be a subset of users and ciu ∈ Cu (1≤ i≥ ku). Let cjp be a subset of pages and cjp ∈ Cp (1≤ j≥ kp). The pair (ciu,cjp) denotes a Co-Cluster of A. By combining the results of user dimensional clustering and page dimensional clustering, ku × kp Co-clusters are obtained. The objective of the paper is to quantify these Co-clusters in different degree using fuzzy membership function. The proposed Fuzzy Co-Clustering algorithm has three phases

### A. First Phase: User Clustering

1. Compute user fuzzy subset of the user associated $A(U,P)_{nxm}$ using equation 3.
2. Compute user similarity matrix of size 'n x n' using fuzzy similarity measure as defined in equation2.
3. Apply K-Means to user similarity matrix and generate $k_u$ user groups.

### B. Second Phase: Page Clustering

1. Compute page fuzzy subset of the user associated $A(U,P)_{nxm}$ using equation 4.
2. Compute page fuzzy similarity matrix of size 'm x m' using equation2.
3. Apply K-Means to page similarity matrix and generate $k_p$ page groups.

### C. Third Phase: Fuzzy Relation Coefficients

1. Combining the results of user dimensional clustering and page dimensional clustering, to obtain $k_u \times k_p$ Co-clusters .
2. Calculate relation coefficients between user cluster and page cluster of each Co-Cluster using equation5 that indicates the distribution of related users' interest over the page clusters.
3. Calculate relation coefficients between user cluster and page cluster of each Co-Cluster using equation6 that shows which user cluster has more interest in that page cluster.

After performing one dimensional clustering on user fuzzy subset and page fuzzy subset, $k_u$ user clusters and $k_p$ page clusters are related and quantified user clusters' interest in the different degree to different page clusters. It reveals the group of related users' interest in the different group of related web pages. The fuzzy relation co-efficient between user cluster and web page cluster is defined in two ways as

$$f(c_i^u, c_j^p) = \frac{\sum_{u_i \in c_i^u} \sum_{p_j \in c_j^p} \text{Hits}(u_i, p_j)}{\sum_{k=1..n} \sum_{p_j \in c_j^p} \text{Hits}(u_i, p_k)} \quad (5)$$

$$f(c_i^u, c_j^p) = \frac{\sum_{u_i \in c_i^u} \sum_{p_j \in c_j^p} \text{Hits}(u_i, p_j)}{\sum_{k=1..m} \sum_{u_i \in c_i^u} \text{Hits}(u_k, p_j)} \quad (6)$$

Equation5 quantifies the each user clusters' interest for different related web page clusters. Equation 6 quantifies the different users' clusters interest for each web page clusters. The interpretation of the fuzzy co-clustering result can be used to improve direct and target marketing strategy and also used to improve the quality of recommendation systems.

## V. EXPERIMENTATION AND RESULTS

Inorder to evaluate performance of the proposed algorithm, experiment is conducted on the benchmark clickstream dataset of MSNBC.com which describes the sequence of page visits of users on 28 September 1999.

### A. Data Preprocessing

Data preprocessing[8] transforms the data into a format that will be more easily and effectively processed for the purpose of the user. The techniques to preprocess data include data cleaning, data integration, data transformation and data reduction. Clickstream records in the MSNBC dataset is converted into matrix format where elements aij of A(U,P) represents the frequency of the user Ui accesses the web page Pj of a web site during a given period of time. During the user session, the user visited web page categories are marked with the frequency of that page accessed and otherwise 0.

### B. Data filtering

Data filtering is the task of extracting only those records of weblog files, which are essential for the analysis, thus reducing significantly data necessary for further processing. In this paper, data filtering aims to filter out the users who have visited less than 9 page categories of web site. Initially there are 989818 users, after this step number of users are reduced to 1720 users.

### C. Results

K-Means algorithm is applied to the resultant user associated matrix of size 1720 X 17 where $k_u$=10 and $k_p$= 3 was fixed to create ten user clusters and three page clusters. Using equation 4 and equation 5, the relations between user clusters and page clusters were quantified as shown in Fuzzy Relation Coefficient Matrix 1and Matrix 2

## VI. FUZZY RELATION CO-EFFICIENT

Table 1 shows which user and page clusters are more related and it indicate the way of users' clusters interest distribution over all pages' clusters. From the Table1, User Cluster $c_2^u$ has more interest in the page cluster $c_3^p$ because that Co-Cluster 's fuzzy relation value is high. Similarly, interested pages for each user cluster can be found easily and efficiently

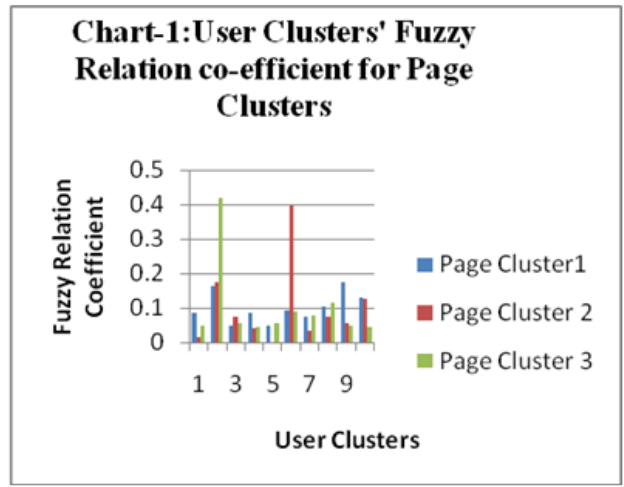

Table 2 shows Co-Cluster's fuzzy relation value by relating user and page clusters. It clearly pictures out which user cluster has more interest for a page cluster. By this way it is easy to identify target related user group for each page cluster and which is useful for target marketing to make recommendations according to their frequent access of web pages during a given period of time.

| Clusters | $c_1^p$ | $c_2^p$ | $c_3^p$ |
|---|---|---|---|
| $c_1^u$ | 0.0839 | 0.0162 | 0.0489 |
| $c_2^u$ | 0.1615 | 0.1739 | 0.4192 |
| $c_3^u$ | 0.0466 | 0.0763 | 0.0574 |
| $c_4^u$ | 0.0847 | 0.0423 | 0.0456 |
| $c_5^u$ | 0.0485 | 0.0019 | 0.0549 |
| $c_6^u$ | 0.0921 | 0.3978 | 0.0877 |
| $c_7^u$ | 0.0739 | 0.0346 | 0.0792 |
| $c_8^u$ | 0.1049 | 0.0752 | 0.1165 |
| $c_9^u$ | 0.1734 | 0.0562 | 0.0475 |
| $c_{10}^u$ | 0.1305 | 0.1256 | 0.0431 |

Table 1 : Users' Cluster Fuzzy Relation Coefficient for Page Clusters

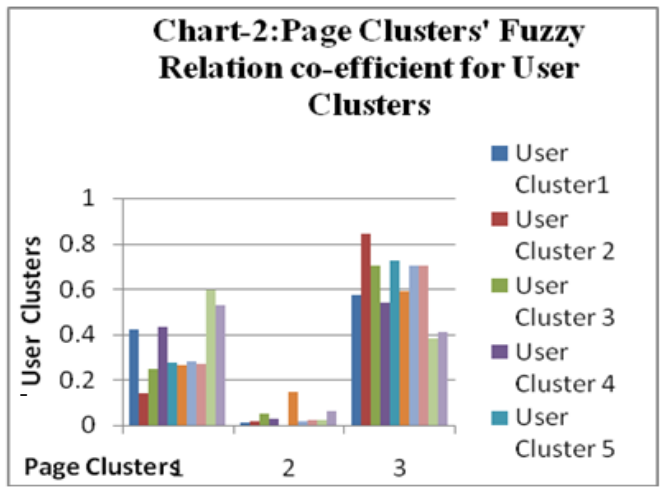

Table 2 shows Co-Cluster's fuzzy relation value by relating user and page clusters. It clearly pictures out which user cluster has more interest for a page cluster. By this way it is easy to identify target related user group for each page cluster and which is useful for target marketing to make recommendations according to their frequent access of web pages during a given period of time.

Interpretation of Co-Cluster result with fuzzy relation value is very helpful to realize how and with which patterns the web site page categories are visited more by the which user cluster. Such information's are useful to the web administrators for web site evaluation or reorganization. Recommend set of related web page category for users group based on the fuzzy relation value also possible.

| Clusters | $c_1^u$ | $c_2^u$ | $c_3^u$ | $c_4^u$ | $c_5^u$ | $c_6^u$ | $c_7^u$ | $c_8^u$ | $c_9^u$ |
|---|---|---|---|---|---|---|---|---|---|
| $c_1^p$ | 0.4192 | 0.1389 | 0.2445 | 0.4309 | 0.274 | 0.2652 | 0.2805 | 0.2712 | 0.595 |
| $c_2^p$ | 0.0103 | 0.0189 | 0.0507 | 0.0273 | 0.0014 | 0.1451 | 0.0167 | 0.0246 | 0.0244 |
| $c_3^p$ | 0.5705 | 0.8422 | 0.7048 | 0.5419 | 0.7246 | 0.5898 | 0.7028 | 0.7041 | 0.3806 |

## VII. CONCLUSION

This paper proposed Fuzzy Co-Clustering algorithm for Cliskstream data and evaluated it with real dataset. The results proved its efficiency in correlating the relevant users and web pages of a web site. Thus, interpretation of Co-cluster results are used by the company for focalized marketing campaigns to an interesting target user cluster. This is a key feature in target marketing. Our Fuzzy Co-Clustering algorithm produces non-overlapping co-clusters. In future it is extended to generate overlapping clusters.